\documentclass[
    aps,prl,twocolumn,
	groupedaddress,superscriptaddress,
	amsfonts,amssymb,amsmath,
	citeautoscript,longbibliography,
	a4paper
	]{revtex4-2}

\usepackage[utf8]{inputenc}
\usepackage[english]{babel}

\usepackage{microtype} 
\usepackage{xspace} 

\usepackage{txfonts}  
\usepackage{txfontsb} 

\usepackage{bm} 

\usepackage{xcolor}
\usepackage[]{graphicx} 
\graphicspath{{figs/}}

\usepackage[]{booktabs}
\usepackage{array}
\usepackage{layouts}
\usepackage{multirow}

\usepackage{enumerate}
\usepackage[inline]{enumitem}

\usepackage{xr}
\makeatletter
\newcommand*{\addFileDependency}[1]{
  \typeout{(#1)}
  \@addtofilelist{#1}
  \IfFileExists{#1}{}{\typeout{No file #1.}}
}
\makeatother

\newcommand*{\myexternaldocument}[1]{%
    \externaldocument{#1}%
    \addFileDependency{#1.tex}%
    \addFileDependency{#1.aux}%
}

\usepackage{hyperref}
\hypersetup{colorlinks,
	linkcolor={blue!75!black!80!yellow},
	citecolor={blue!75!black!80!yellow},
	urlcolor={blue!75!black!80!yellow}
}

\usepackage[capitalize,nameinlink]{cleveref}

\crefname{subequations}{Eqs.}{Eqs.} 
\Crefname{subequations}{Eqs.}{Eqs.}
\crefformat{subequations}{#2Eqs.~(#1)#3}
\Crefformat{subequations}{#2Eqs.~(#1)#3}
\crefname{page}{p.}{p.} 

\usepackage{placeins}

\usepackage{siunitx}
\DeclareSIUnit[number-unit-product = ]\percent{\char`\%} 

\usepackage[centering,hmargin=18mm,tmargin=24mm,bmargin=24mm]{geometry}

\usepackage{soul}

\usepackage{textcomp} 
\usepackage{xifthen}
\usepackage{etoolbox}
\newboolean{togglecomments}
\newboolean{toggletodos}
\newboolean{togglechanges}

\setboolean{togglecomments}{true}
\setboolean{toggletodos}{true}
\setboolean{togglechanges}{false} 

\newcommand{\textblacksquare}{$\blacksquare$}
\newcommand{\todo}[1]{\ifbool{toggletodos}%
	{\textcolor{green!65!black}{\small\textsf{{}\textsuperscript{\textsc{\textsf{todo}}}}[\ignorespaces#1]}} 
	{}}     
\newcommand{\comment}[2]{\ifbool{togglecomments}%
		{\textcolor{blue!70!black}{\small\sf\textsuperscript{\textsc{\textsf{\ignorespaces#1}}}[\ignorespaces#2]}} 
		{}}     
\newcommand{\swap}[2]{\ifbool{togglechanges}
	{\ignorespaces#2}  
	{\textcolor{red!80!black}{[\ignorespaces#1]}\textrightarrow{}\textcolor{green!65!black}{[\ignorespaces#2]}}}
\newcommand{\remove}[1]{\ifbool{togglechanges}
	{}    
	{\textcolor{red!80!black}{\ignorespaces#1}}}
\newcommand{\inset}[1]{\ifbool{togglechanges}
	{\ignorespaces#1}  
	{\textcolor{green!65!black}{\ignorespaces#1}}}

\newcommand{\citeremind}[1]{%
	[\textcolor{blue!75!black!80!yellow}{\textblacksquare%
		\ifthenelse{\isempty{#1}}{}{\textsuperscript{\tiny\textsf{\ignorespaces#1}}}%
	}]\xspace}

\newcommand{\kv}{\mathbf{k}}

\newcommand{\tsc}[1]{\text{\textsc{#1}}}
\newcommand{\e}{\mathrm{e}}

\newcommand{\iu}{\mathrm{i}}

\newcommand{\ie}{i.e.,\@\xspace} 

\newcommand{\eg}{e.g.,\@\xspace}

\newcommand{\appropto}{\mathrel{\vcenter{
			\offinterlineskip\halign{\hfil$##$\cr
				\propto\cr\noalign{\kern.2pt}\sim\cr\noalign{\kern-2.5pt}}}}}

\let\Im\relax 
\DeclareMathOperator{\Im}{Im}

\newcommand{\suppsec}{Supplemental Section\xspace}

\DeclareFontFamily{U}{mathx}{\hyphenchar\font45}
\DeclareFontShape{U}{mathx}{m}{n}{<5> <6> <7> <8> <9> <10>
                                  <10.95> <12> <14.4> <17.28> <20.74> <24.88>
                                  mathx10}{}
\DeclareSymbolFont{mathx}{U}{mathx}{m}{n}
\DeclareFontSubstitution{U}{mathx}{m}{n}

\makeatletter
\newcommand{\raisemath}[1]{\mathpalette{\raisem@th{#1}}}
\newcommand{\raisem@th}[3]{\raisebox{#1}{$#2#3$}}
\makeatother

\renewcommand{\paragraph}[1]{\vskip .5ex\noindent\emph{#1.}---\ignorespaces}

\usepackage[eulergreek]{sansmath}
\makeatletter
\renewcommand\@make@capt@title[2]{%
    \@ifx@empty\float@link{\@firstofone}{\expandafter\href\expandafter{\float@link}}%
    \sisetup{math-sf=\textsf}%
    \sansmath\sffamily\textbf{#1\@caption@fignum@sep}#2 
}%

\makeatother

\myexternaldocument{supp}
\usepackage{physics} 
\usepackage{soul} 
\setstcolor{red}

\graphicspath{{figures/}}

\setboolean{togglechanges}{false} 
\setboolean{togglecomments}{true} 

\begin{document}

\title{Quasicrystalline Weyl points and dense Fermi--Bragg arcs}

\newcommand{\mitaffil}{Department of Physics, Massachusetts Institute of Technology, Cambridge, Massachusetts 02139, USA}
\newcommand{\dtuaffil}{Department of Photonics and Electrical Engineering, Technical University of Denmark, Lyngby 2800, Denmark}

\author{Andr\'e Grossi e Fonseca}
\email{agfons@mit.edu}
\affiliation{\mitaffil}
\author{Thomas Christensen}
\email{thomas@dtu.dk}
\affiliation{\mitaffil}
\affiliation{\dtuaffil}
\author{John D.~Joannopoulos}
\affiliation{\mitaffil}
\author{Marin Solja\v{c}i\'c}
\affiliation{\mitaffil}

\begin{abstract}
    We introduce a general mechanism for obtaining Weyl points in a stack of 2D quasicrystals, which can be extended to any stack of aperiodic layers.
    We do so by driving a topological phase transition with the vertical crystal-momentum as the tuning parameter, which leads to gap closures at the critical points sourcing Berry curvature.
    To illustrate, we use a simple 3D generalization of the Qi--Wu--Zhang model defined on a Penrose quasicrystal.
    The presence of Weyl points is established via the local Chern marker, projected band structure and density of states. Interestingly, we uncover an analogue of Fermi arcs in the quasicrystalline setting, which we deem Fermi--Bragg arcs, densely distributed lines connecting the band degeneracies and indexed by the Bragg peaks.
    Signatures of such surface states in quantum oscillations and the prospect of a fully quasicrystalline Weyl system are also discussed.
    The flexibility of our proposal brings new opportunities for realizing other gapless topological phases in aperiodic systems, paving the way for a significantly expanded role for topological band theory.
\end{abstract}
\maketitle

The past decades have seen the development of a broad understanding of band theory from topological perspectives in both quantum~\cite{KaneHasan2010, Cooper2019} and classical~\cite{Lu2014, TopPhot2019, Ma2019} crystalline systems.
This understanding includes both gapped phases such as Chern insulators~\cite{Haldane:1988}, topological insulators~\cite{Fu:2007,Kane:2005}, and topological crystalline insulators~\cite{Fu:2011}, as well as gapless phases including Dirac~\cite{CastroNeto2009} and Weyl~\cite{Armitage2018} points, and nodal lines~\cite{Burkov2011}.
With the recent introduction of several real-space formulations of topological invariants~\cite{Bianco2011, Loring2011, Kitaev2006, Cerjan2022}, the assumption of lattice periodicity has been relaxed for several classes of gapped~\cite{Huang2022, Agarwala2017, Chen2020} and metallic~\cite{Yang2019} topological phases.
In this Letter, we show that gapless topological phases also do not fundamentally rely on crystallinity by direct construction of a model with spin-1/2 Weyl points in a stack of quasicrystals.

Weyl points represent perhaps the simplest three-dimensional (3D) gapless topological phase. Introduced by Hermann Weyl in 1929~\cite{Weyl:1929} as the massless solutions of the Dirac equation, they were experimentally realized nearly 80 years later in TaAs~\cite{Xu2015, Lv2015} and double gyroid photonic crystals~\cite{Lu2015}.
In crystalline settings, they correspond to solutions of the Weyl Hamiltonian $H(\kv) = v_x k_x \sigma_x + v_y k_y \sigma_y + v_z k_z \sigma_z$, with $k_i$, $v_i$, and $\sigma_i$ denoting the components of crystal momentum, group velocity, and Pauli matrix vector, respectively, and exhibit a gapless linear dispersion.
Their associated Fermi-arc surface states~\cite{Wan2011}, connection to high-energy physics~\cite{Volovik2003}, and unique optical properties~\cite{Zhou:2017,Fan2022} have fueled intense interest.
More fundamentally, Weyl points are key conceptual building blocks of many topological phenomena, since they act as monopole sources and sinks of Berry curvature~\cite{Vanderbilt2018, Armitage2018}; we take this to be their defining attribute throughout this Letter.

\begin{figure}[!htb]
    \centering
    \includegraphics[scale=1]{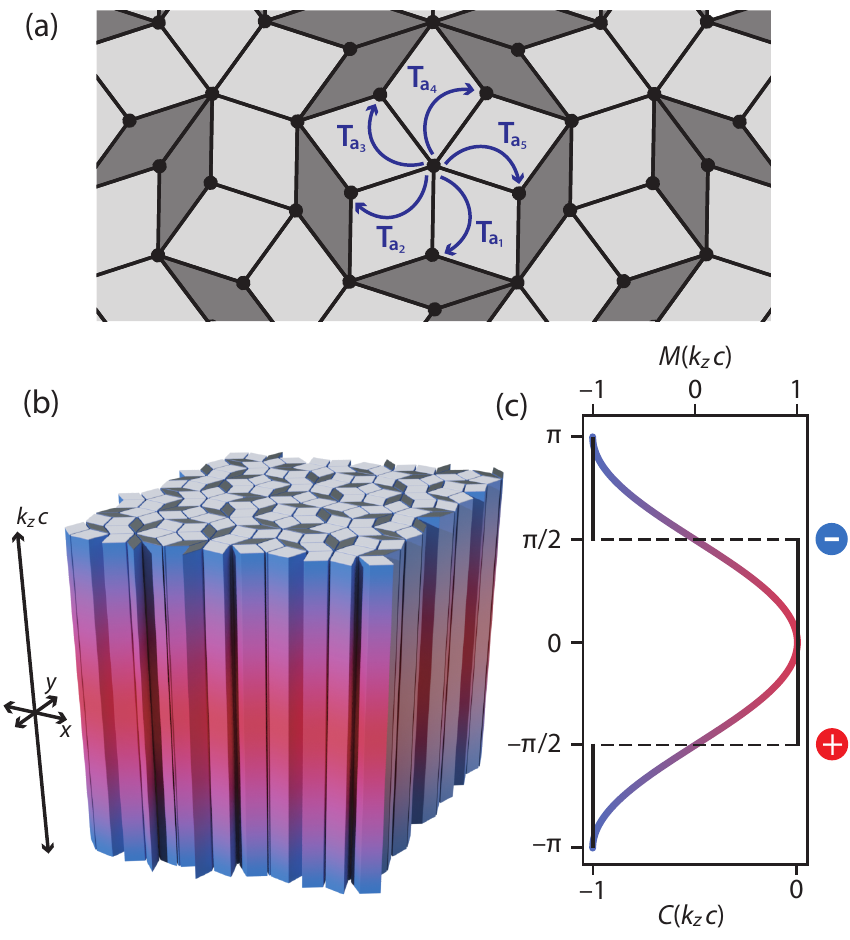}
    \caption{%
        \textbf{Mechanism for Weyl points in a quasicrystal stack.}
        (a)~Top view of a small section of a Penrose quasicrystal, with fat and skinny rhombi in light and dark grey, respectively.
        Excitations hop between vertices along the edges of adjacent rhombi with
        amplitudes $\mathbf{T}_{\mathbf{a}}$  given by \cref{eq:QWZ}.
        The possible hopping vectors are $\mathbf{a}_j = (\sin 2\pi j/5, \cos 2\pi j/5)a, \, j=1,\ldots,5$ with bond length $a$.
        (b)~3D generalization of (a), obtained by sinusoidally varying the mass term $M$ of \cref{eq:QWZ} over the out-of-plane crystal momentum $k_z$ with real-space period $c$, as indicated by the color gradient.
        (c)~Mass parameter $M$ (with $M_0=1$) and expected Chern number $C$ as a function of $k_z$.
        $C$ changes by $\pm1$ when $M$ changes sign, indicating the presence of a Weyl point with chirality $\pm 1$ (circles).%
 }
    \label{fig:fig1}
\end{figure}

Nonetheless, the understanding and realization of Weyl points remains firmly rooted in $\kv$-space centered perspectives, lacking generalizations to systems where Bloch's theorem does not apply.
Here, we introduce a general mechanism for realizing Weyl physics in a stack of quasicrystalline Hamiltonians and discuss their associated bulk and boundary signatures.
Our approach, generalizable to any family of aperiodic Hamiltonians, relies on a simple idea, borrowed from the crystalline setting:
any two distinct two-dimensional (2D) manifolds separated by a single Weyl point must be characterized by Chern numbers differing by the enclosed Weyl point chirality~\cite{Vanderbilt2018}.
We exploit this fact by driving a band inversion which changes the Chern number of a quasicrystalline bulk 2D Hamiltonian. We take the parameter controlling the topological transition to be crystal momentum $k_z$ associated with stacking in the $z$-direction, which leads to the emergence of Weyl points at the inversion points (\cref{fig:fig1}).
We illustrate our idea in a simple tight-binding Hamiltonian defined on vertices of a Penrose tiling~\cite{Penrose1974} with a $k_z$-varying mass term, which captures the essential features of our proposal.
The bulk signatures of the associated Weyl points are consistent with those of the crystalline setting, as we establish explicitly via the local Chern marker~\cite{Bianco2011}, bulk energy dispersion, and density of states.
Interestingly, we uncover a quasicrystalline version of Fermi arcs which we call Fermi--Bragg arcs, whose characteristics sharply distinguish them from their crystalline counterparts: they are densely distributed~\footnote{We employ the word ``dense'' in the mathematical sense, \ie a set is dense if in between any two elements one can always find at least one other element.} in Fourier space and terminate at vertical lines through the Bragg peaks of the diffraction spectrum.

\paragraph{Model and topology} 
Our construction begins with a 2D quasicrystal.
Concretely, we consider a tight-binding model defined on the vertices of a Penrose tiling~\cite{Penrose1974}, a small section of which is shown in \cref{fig:fig1}(a).
Because we will consider both open and periodic boundary conditions (OBC and PBC), we construct the quasicrystal through Penrose tiling approximants~\cite{Bruijn198}, finite supercells of increasing order $n$ that systematically approximate the Penrose quasicrystal as $n \rightarrow \infty$ while restricting the number of defects to two per unit cell (corresponding to edges along which the matching rules are violated)~\cite{Wohlman1988}.
The number of vertices at order $n$ is $F_{2n+3} + F_{2n+5}$, where $F_n$ is the $n$th Fibonacci number~\cite{Tsunetsugu1991} (\suppsec~S1).

We couple the vertices of this lattice via a generalized Qi--Wu--Zhang (QWZ) model~\cite{QWZ2006} [\cref{fig:fig1}(a)], whose two-band phase diagram exhibits Chern phases.
The real-space tight-binding Hamiltonian takes the form~\cite{ZhenRong:2022}:
\begin{subequations}\label{eq:QWZ}
\begin{align}
    &H_{\tsc{qwz}}(M)
    = 
    t\sum_{\mathbf{R}}
    \left[
    c^{\dag}_{\mathbf{R}}\boldsymbol{\epsilon}(M)c_{\mathbf{R}} +
    \sum_{\mathbf{a}} c^{\dag}_{\mathbf{R}}\mathbf{T}_{\mathbf{a}}c_{\mathbf{R}+\mathbf{a}}\right],
    \\
    &\boldsymbol{\epsilon} (M) = (2+M) \sigma_z, \quad
    \mathbf{T}_\mathbf{a} = 
    \frac{1}{2}\begin{pmatrix}
    1 & \iu e^{-\iu \theta_{\mathbf{a}}}\\
    \iu  e^{\iu \theta_{\mathbf{a}}} & -1
    \end{pmatrix},
\end{align}
\end{subequations}
where $c^{\dag}_{\mathbf{R}} = (c^{\dag}_{\mathbf{R},1}, c^{\dag}_{\mathbf{R},2})$, with $c^{\dag}_{\mathbf{R},\alpha}$ creating an excitation at site $\mathbf{R}$ and site degree of freedom $\alpha$, $t$ is the overall hopping amplitude and the sum runs over all vertex positions $\mathbf{R}$ of the quasicrystal and neighbors connected to $\mathbf{R}$ by a rhombus edge with vector $\mathbf{a}$.
Finally, $\theta_{\mathbf{a}}$ is the angle of the edge $\mathbf{a}$ with the $x$-axis and $M$ is a mass parameter controlling the topological phase, corresponding physically to an on-site potential staggering.
Note that the Hamiltonian is Hermitian, although the hopping matrices $\mathbf{T}_{\mathbf{a}}$ are not.
Concretely, the QWZ model can be realized \eg with spin-orbit coupled $s$-orbitals and ferromagnetic moments as the site degrees of freedom~\cite{Qi2011}.
On a square lattice, it undergoes topological phase transitions at $M = -4, -2$ and $0$ (\suppsec~S4.A).
In particular, the Chern number $C$ of the valence band changes from $-1$ to $0$ when $M$ changes sign from negative to positive. 
The choice of the QWZ model is chiefly motivated by simplicity: our main qualitative conclusions only require certain topological phase transitions.

To compute the topological phase diagram in our quasicrystalline setting, we require a notion of Chern numbers that does not rely on crystal-momentum.
We adopt the formalism of the local Chern marker~\cite{Bianco2011}, which defines a Chern amplitude $\mathcal{C}_i$ at each lattice site $i$ with coordinates $(x_i,y_i)$ as~\cite{Tran2015}:%
\begin{subequations}
\begin{equation}
    \label{LCM}
    \mathcal{C}_i = -\frac{4\pi}{\Omega} \Im \bigg[ \sum_j \bra{i}x_\mathcal{Q}\ket{j}\bra{j}y_\mathcal{P}\ket{i}\bigg], 
\end{equation}
where
\begin{alignat}{2}
    &\bra{i}x_\mathcal{Q}\ket{j}=\sum_k \mathcal{Q}_{ik}x_k\mathcal{P}_{kj},
    \quad
    &&\bra{j}y_\mathcal{P}\ket{i} = \sum_k \mathcal{P}_{jk}y_k\mathcal{Q}_{ki},
    \\
    &\quad\ 
    \mathcal{P} = \sum_{\varepsilon_n < \varepsilon_{\tsc{F}}} \ket{\psi_n}\bra{\psi_n},
    \quad
    &&\qquad\ \
    \mathcal{Q} = \mathbb{1}-\mathcal{P},
\end{alignat}
\end{subequations}
with $\Omega$ denoting the unit cell area, which for the Penrose quasicrystal we take to be the weighted average of the fat and skinny rhombus areas (\suppsec~S2) and $\ket{\psi_n}$ an eigenstate of index $n$ and energy $\varepsilon_n$. 

For a finite system, the local Chern marker averaged over all lattice sites vanishes~\cite{Bianco2011}.
The average over a finite bulk region, however, is well-quantized and converges to the conventional Chern number in the thermodynamic limit~\cite{Tran2015}.
For concreteness, we define this average over a disk $D$ of radius $R$ containing $N$ sites:%
\begin{equation}
\label{aLCM}
    \mathcal{C}_D = \frac{1}{N}\sum_{i \in D}\mathcal{C}_i.
\end{equation}
In practice, we choose $R$ large enough to avoid microscopic fluctuations but small enough to be restricted to the bulk  (\suppsec~S2).
Similarly to the square-lattice setting, we observe that the averaged Chern marker of the valence band in the quasicrystalline QWZ model, within small fluctuations, is quantized to $-1$ for $M<0$ and to $0$ for $M>0$, changing value near $M=0$.

Given these observations, we extend the 2D quasicrystalline QWZ model in a third, periodic dimension ($z$) of period $c$.
Specifically, we imagine stacking 2D quasicrystal layers along $z$ in such a manner that the mass varies sinusoidally in the crystal momentum $k_z$, as illustrated in \cref{fig:fig1}(c,d).
This leads to a simple 3D extension of the QWZ Hamiltonian:
\begin{equation}
\label{iQWZ}
    H(k_z) = H_{\tsc{qwz}}(M_0 \cos k_z c).
\end{equation}
The mass amplitude $M_0$ controls the number of phase transitions in $k_z$ (\suppsec~S4): for simplicity, we take $M_0=1$ going forward.
The model has neither inversion nor time-reversal symmetry---the latter is broken by the off-diagonal elements of $\mathbf{T}_{\mathbf{a}}$---but the out-of-plane mirror symmetry leads to Weyl point pairs at $\pm k_z$.
In addition, sublattice symmetry and the two-band nature of the model pin all Weyl points to zero energy.
The expected phase diagram---the evolution of the Chern number $C(k_z)$ with $k_z$---is shown schematically in \cref{fig:fig1}(c).
As we will show, the associated phase transition points $k_z^\pm \approx \pm \pi/2c$ are accompanied by band closings, each corresponding to a Weyl point of chirality $\chi^\pm = \lim_{\delta\rightarrow 0^+}C(k_z^\pm + \delta) - C(k_z^\pm - \delta) = \mp 1$.

\begin{figure}
    \centering
    \includegraphics[scale=1]{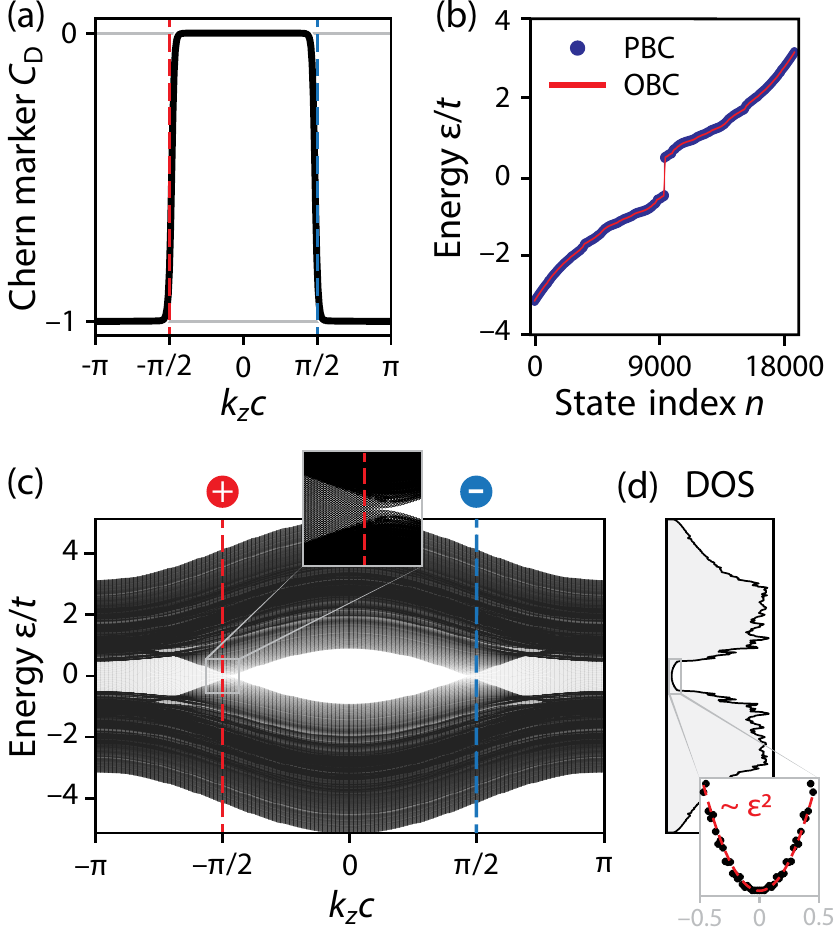}
    \caption{%
        \textbf{Bulk signatures of quasicrystalline Weyl points.}
        (a)~Disk-averaged Chern marker $\mathcal{C}_D$ for the valence band of $k_z$-slices of $H(k_z)$ (disk radius $R=12a$).
        Topological phase transitions occur at $k_z c \approx \pm \pi/2$.
        (b)~The energy spectrum $\varepsilon_n(k_z)$ of $H(k_z)$ at $k_z c = \pi$, as a function of state index $n$.
        While the spectrum is gapped under PBCs, a band of in-gap chiral edge states appears under OBCs, implying a nontrivial Chern number.
        (c)~Projected band structure under OBCs as a function of $k_z$.
        For $|k_z c| \lesssim \pi/2$, the system is gapped, whereas for $|k_z c| \gtrsim \pi/2$ edge states populate the bulk gap, as shown in the inset.
        Red dashed lines indicate the gap-closing points for a square lattice.
        (d)~DOS under PBCs, displaying a quadratic energy-dependence near zero energy (inset: quadratic fit in red dashed line).
        (a) and (b--d) were computed from Penrose tiling approximants of order $n=6$ (3571 sites) and $n=7$ (9349 sites), respectively.}
    \label{fig:fig2}
\end{figure}

\paragraph{Bulk signatures}
We now consider the bulk features of our 3D model in detail by numerically solving \cref{iQWZ} for large Penrose quasicrystal tilings (\cref{fig:fig2}), identifying two well-known bulk signatures of Weyl points.
First, the disk-averaged local Chern marker $\mathcal{C}_D(k_z)$ [\cref{fig:fig2}(a)] exhibits sharp topological phase transitions at $k_zc \approx \pi/2$ where the mass parameter $M(k_z)$ changes sign, in excellent agreement with our qualitative expectations.
Interestingly, the phase transition point at $M \approx 0$ agrees with that of the square lattice, suggesting that this aspect is largely independent of lattice details.
The valence band of $H(k_z)$ has $\mathcal{C}_D = -1$ for $|k_z c| \gtrsim \pi/2$, and is otherwise trivial.
\Cref{fig:fig2}(b) shows the spectrum of $H(k_z)$ at $k_zc=\pi$.
Under PBCs, the spectrum shows a clear bulk gap; under OBCs, this gap is bridged by chiral edge modes, consistent with a Chern phase.\\
\indent We also computed the full spectrum of $H(k_z)$ as a function of $k_z$ under OBCs, shown in \cref{fig:fig2}(c).
The existence of a small but finite gap is due to finite-size effects, vanishing in the limit of increasing quasicrystal tiling order $n\rightarrow \infty$ (\suppsec~S3). In this limit, the spectrum therefore hosts a linear gapless dispersion near the degeneracy points, which constitutes the second bulk signature.
Finally, since both trivial and nontrivial $k_z$ slices of the spectrum share an overlapping gap, the band crossings are well-isolated from other bulk states and their density of states (DOS) can consequently be probed in isolation.
This allows for the identification of a third bulk signature of Weyl points: their well-known form $\mathrm{DOS}(\varepsilon) \propto \varepsilon^2$ under PBCs [\cref{fig:fig2}(d)].

\paragraph{Surface signatures and Fermi--Bragg arcs}
A hallmark of Weyl points in crystals is the presence of anomalous surface states deemed Fermi arcs. A consequence of the bulk-boundary correspondence, they are arc-like Fermi surfaces that connect the projections of Weyl points on the surface Brillouin zone~\cite{Wan2011}. 

\begin{figure}[h!]
    \centering
    \includegraphics[scale=1]{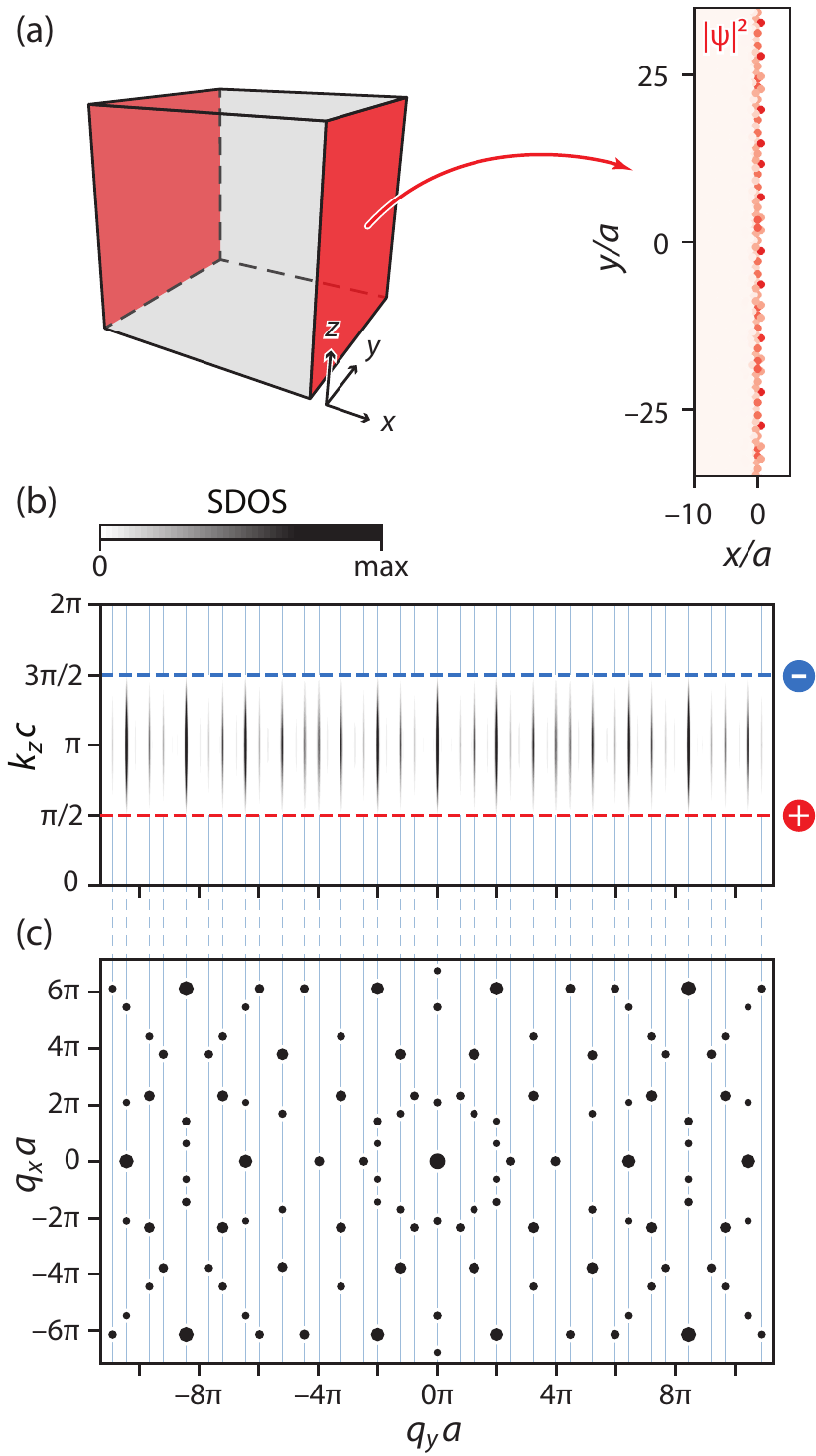}
    \caption{%
        \textbf{Surface signatures of Weyl points in real and momentum space.}
        (a)~Upon imposing OBCs in both $x$ and $y$ and zooming into the $x$ boundary, we observe localized edge modes for values of $k_z$ in the topological regime (edge state shown for $k_z c = \pi$ on a linear color scale).
        (b)~Spectral DOS at the $x$ boundaries. Horizontal red dashed lines bound the topological region $\pi/2 < k_z c < 3\pi/2$, showing that arcs connect Weyl points of opposite chirality. (c)~Structure factor in Fourier space (only brightest spots are shown and spot size is proportional to intensity).
        Vertical blue lines show that $q_y$ projections of Bragg peaks precisely line up with Fermi arcs.
        (a--c) were computed from Penrose tiling approximants of order $n = 6$ (3571 sites).
        }
    \label{fig:fig3}
\end{figure}

In our quasicrystalline setting, with only periodicity in $z$, we lack a simple surface BZ equivalent.
Instead, we imagine an angle-resolved diffraction measurement on an exposed surface of our quasicrystal stack, say, a $(100)$ termination spanning the $(y,z)$ plane, resolved in crystal momentum $k_z$ along $z$ and in Fourier momentum $q_y$ along $y$ [\cref{fig:fig3}(a)].
Lacking periodicity in $y$, $q_y$ is unambiguously defined in the infinite domain $(-\infty,\infty)$, rather than the finite domain $[0,2\pi/c)$ of $k_z$.
For a $y$-periodic system, $q_y$ simply maps to infinite adjacent copies of the BZ along $y$ in an extended zone scheme (\suppsec~S7.A).
We may then evaluate the spectral density of states $\mathop{\mathrm{SDOS}}(x, q_y, k_z; \varepsilon)$---\ie the $x$-position-, $q_y$-Fourier-component-, $k_z$-crystal-momentum-resolved DOS.
Physically, $\mathop{\mathrm{SDOS}}(x, q_y, k_z; \varepsilon)$ dictates the DOS to which an incident plane wave ${\sim}\,\e^{\iu k_z z}\e^{\iu q_y y}$ of energy $\varepsilon$ can couple at position at $x$.
\Cref{fig:fig3}(b) shows the spectral DOS for our quasicrystal stack, evaluated at the $x$-boundary of a $(100)$ surface termination and at $\varepsilon=0$ (where bulk contributions vanish).
A rich structure of vertical arcs of varying intensities appears in $q_y$, all connecting the projections of Weyl points at $\approx \pm \pi/2c$ across $k_z$.

To interpret the appearance of such arcs, we first draw an analogy with the crystalline square-lattice case.
When considered in an extended zone scheme, a conventional crystal with Weyl points also exhibits multiple Fermi arcs, with each arc associated with a copy of the first BZ, generated by a shift by a reciprocal lattice vector (\suppsec~S7.A).
Each such Fermi arc copy can consequently be labelled by a 2-dimensional integer vector, \ie the Miller indices.
Equivalently, each BZ- and arc-copy is associated with a specific Bragg peak of the structure factor $S(\mathbf{q}) \propto \big|\sum_{\mathbf{R}} \e^{\iu \mathbf{q}\cdot \mathbf{R}}\big|^2$ (lattice sites $\mathbf{R}$).
Motivated by this, we plot $S(\mathbf{q})$ for the Penrose tiling in \cref{fig:fig3}(c).
We see perfect alignment of the spectral arcs with the Bragg peaks' $q_y$ positions---as well as some correlation of spectral DOS and structure factor intensities.
Due to this correspondence, we call the associated surface states Fermi--Bragg arcs.
We note that the arcs' verticality in $k_z$ is due to the simplicity of our tight-binding model; for more general hoppings, the arcs curve---but their terminations remain pinned to the Bragg peaks (\suppsec~S7.A).
Each of these arcs can be labelled by a family of generalized Miller indices, each a $5$-dimensional integer vector stemming from the 5D hypercube from which the Penrose quasicrystal is generated (\suppsec~S6).
In this perspective, the Fermi--Bragg arcs are analogous to conventional Fermi arcs, but now densely distributed rather than periodically, a feature inherited from the dense structure factor of quasicrystals~\cite{Levine1986}.
Furthermore, the absence of translational symmetry implies that arcs will generically have different lengths in momentum space, leading to a series of incommensurate peaks in the Fourier transform of a Shubnikov--de Haas measurement (\suppsec~S7.B).

\paragraph{Real-space model}
The Hamiltonian of our quasicrystal stack in \cref{iQWZ} is in a mixed-coordinate form, with in-plane (quasicrystalline) coupling defined in $(x,y)$-space and out-of-plane (periodic) coupling defined in $k_z$-space.
To understand how the model can be realized in practice, we transform the $k_z$-dependence to a $z$-dependence by an inverse Fourier transform (\suppsec~S5), and find that it is equivalent to a stack of quasicrystal layers tuned to the critical $M=0$ point with diagonal interlayer hopping (\ie between identical layer degrees of freedom).
However, neither fine-tuning of $M$ nor the restriction to diagonal hopping is strictly necessary (\suppsec~S5.B): off-diagonal mixing and small detunings away from $M = 0$ simply shifts the Weyl points in $k_z$.
The simplicity of our proposal---\ie stacking of layers near a Dirac point phase-transition in addition to a time-reversal-breaking---indicates that the specific details of the QWZ couplings and Penrose tiling may be inessential to the relevant physics, and suggests significant leeway in experimental implementations.
Opportunities in an electronic setting include magnetic decagonal quasicrystals~\cite{Yokoyama1997}, stacks of magnetic thin films~\cite{Dorini2021}, or magnetically-doped layers~\cite{Chang2013}.
We also highlight classical implementations, particularly in photonic platforms, exploiting gyromagnetic~\cite{Wang2009, Zhang2023} or gyroelectric~\cite{Qian2023} responses, but also more broadly, \eg in mechanics~\cite{Nash2015} or acoustics~\cite{Ding:2019}.
This simplicity may also guide the way to time-reversal-invariant but inversion-broken realizations of quasicrystalline Weyl points~\cite{Cain2020, Timusk2013, Bendersky1985}.

\paragraph{Discussion and outlook}
In this Letter, we have proposed a general mechanism for obtaining a gapless topological phase---Weyl points specifically---in a noncrystalline setting by driving a 2D quasicrystalline Hamiltonian through a topological phase transition.
To illustrate the mechanism, we introduced a simple 3D generalization of the QWZ model on a Penrose quasicrystal stack.
The model hosts pairs of Weyl points, as we verify by identifying several distinct signatures.
From the bulk perspective, we evaluate the phase diagram of the local Chern marker and find topological phase transitions with associated linear energy dispersion and quadratic DOS dependence, all consistent with the existence of bulk Weyl points.
From the surface perspective, we find an analogue of Fermi arcs, which we call Fermi--Braggs arcs, characterized by densely distributed arc-like structures terminating at the Bragg peaks of the diffraction spectrum.
These characteristics are expected to be universal for Weyl points in quasicrystals, and pose an exciting opportunity for experimental observation. Furthermore, the simplicity of our model in real-space suggests that quasicrystalline Weyl points could exist in much more general settings; the recent observations of 3D band crossings in a stack of Ta\textsubscript{1.6}Te quasicrystal layers~\cite{Cain2020} and the proposed explanation of optical conductivity features in 3D quasicrystals in terms of a Weyl phase~\cite{Timusk2013} are interesting outlooks to this.

An important remaining question is whether a \emph{fully} quasicrystalline system can host Weyl points, \ie whether the out-of-plane periodicity can be replaced by a quasicrystalline dimension as well.
We argue that it can and is possible to realize with outset in our model:
for example, we may imagine adiabatically breaking periodicity along $z$ by introducing a quasicrystalline modulation, \eg an Aubry--Andr\'e potential~\cite{AA1980}.
Even though a generic aperiodic weak perturbation leads to a finite, albeit exponentially small, density of states at crystalline Weyl points~\cite{Nandkishore2014, Pixley2016}, it has been shown both numerically~\cite{Pixley2018} and analytically~\cite{Mastropietro2020} that the semimetallic phase is robust to modest quasiperiodic perturbations.
Meanwhile, certain bulk signatures, \eg the DOS, should remain unaffected, since there are no bulk bands that can obscure the Weyl points, assuming sufficiently weak potentials. 
In this scenario, we expect the Fermi--Bragg arcs to be even richer and densely distributed across all surface diffraction planes.
Although $k_z$ is no longer well-defined in a fully aperiodic setting, real-space methods such as the spectral localizer~\cite{Loring2017, Schulz2021} would remain suitable to track the number of Berry monopoles.

An exciting outlook is to use a non-periodic mass term, realized \eg through a synthetic dimension, which would allow simultaneously breaking $z$-periodicity and potential violations of the Nielsen--Ninomiya theorem \cite{Nielsen1981.1, Nielsen1981.2}.
More broadly, our recipe for realizing gapless topology in a quasicrystalline setting may serve as a starting point for more exotic topological phases, such as those protected by noncrystallographic point group symmetries or endowed with higher topological charges or degeneracies, thereby complementing the zoo of crystalline topological quasiparticles~\cite{Bradlyn2016, Tang2021, Yu2022}.

\vskip 2ex
The authors thank Sachin Vaidya and Aidan Reddy for stimulating discussions and helpful suggestions.
A.G.F.\ acknowledges support from the Henry W. Kendall Fellowship and the Whiteman Fellowship.
T.C.\ acknowledges the support of a research grant (project no.~42106) from Villum Fonden.
This research is based upon work supported in part by the Air Force Office of Scientific Research under Grant No.\ FA9550-20-1-0115 and FA9550-21-1-0299, the U.S.\ Office of Naval Research (ONR) Multidisciplinary University Research Initiative (MURI) Grant No.\ N00014-20-1-2325 on Robust Photonic Materials with High-Order Topological Protection, and the U.S.\ Army Research Office through the Institute for Soldier Nanotechnologies at MIT under Collaborative Agreement No.\ W911NF-18-2-0048.
The MIT SuperCloud and Lincoln Laboratory Supercomputing Center provided computing resources that contributed to the results reported in this work.


%

\end{document}